\documentclass[twocolumn,amsmath,amssymb,superscriptaddress,prb]{revtex4}

\usepackage{graphicx}
\usepackage{amsmath,amssymb,amsthm}

\newcommand{\beq}{\begin{equation}}
\newcommand{\eeq}{\end{equation}}
\newcommand{\beqa}{\begin{eqnarray}}
\newcommand{\eeqa}{\end{eqnarray}}
\newcommand{\e}{\mathrm{e}}
\newcommand{\w}{\omega}

\newcommand{\ket}[1]{\left| #1 \right\rangle}

\newcommand{\ketbra}[2]{\left|#1\right\rangle\hskip-1mm\left\langle #2\right|}

\newcommand{\kv}{ {\bf k} }

\begin{document}

\title{A general approach to quantum dynamics using a variational master equation: Application to phonon-damped Rabi rotations in quantum dots}
\author{Dara P. S. McCutcheon} \email{dara.mccutcheon@ucl.ac.uk}
\affiliation{Department of Physics and Astronomy, University College London, Gower Street, London WC1E 6BT, UK}
\affiliation{London Centre for Nanotechnology, University College London}
\author{Nikesh S. Dattani}
\affiliation{Department of Materials, Oxford University, Oxford OX1 3PH, UK}
\author{Erik M. Gauger}
\affiliation{Department of Materials, Oxford University, Oxford OX1 3PH, UK}
\author{Brendon W. Lovett}
\affiliation{School of Engineering and Physical Sciences, Heriot Watt University, Edinburgh EH14 4AS, UK}
\affiliation{Department of Materials, Oxford University, Oxford OX1 3PH, UK}
\author{Ahsan Nazir}\email{a.nazir@imperial.ac.uk}
\affiliation{Blackett Laboratory, Imperial College London, London SW7 2AZ, UK}
\affiliation{Department of Physics and Astronomy, University College London, Gower Street, London WC1E 6BT, UK}

\date{\today}

\begin{abstract}

We develop a versatile master equation approach to describe the non-equilibrium dynamics of a two-level system in contact with a bosonic environment, which allows for the exploration of a wide range of parameter regimes within a single formalism. As an experimentally relevant example, we apply this technique to the study of excitonic Rabi rotations in a driven quantum dot, and compare its predictions to the numerical Feynman integral approach. 
We find excellent agreement between the two methods across a generally difficult range of parameters. 
In particular, the variational master equation technique 
captures effects usually considered to be non-perturbative, such as multi-phonon processes and bath-induced driving renormalisation, and can give reliable results even in regimes in which 
previous master equation approaches fail.

\end{abstract}

\maketitle

{\it Introduction} - 
The recent experimental characterisation of exciton-phonon interactions in a coherently-driven semiconductor quantum dot (QD), and their interpretation in terms of a two-level system in contact with a bosonic environment,~\cite{ramsay10, ramsay10_2} have demonstrated that QDs offer a natural platform in which to explore dissipative dynamics in the solid-state. In particular, the interplay between laser-driven coherent excitonic oscillations and incoherent phonon-induced processes can lead to a rich dynamical behaviour.~\cite{ramsay10, ramsay10_2, vagov2007} For example, the phenomenon of {\it excitation-induced dephasing} has been the subject of intense experimental and theoretical investigation recently.~\cite{ramsay10, ramsay10_2, ulrich11, roy11, vagov2007, machnikowski04,nazir08, forstner03, mccutcheon10_2,krugel05, mogilevtsev08} These studies reveal that the driven exciton oscillation frequency and damping rate depend sensitively, and non-monotonically, on the energy scales of both the driven QD and the bulk phonon modes. However, the relatively large range of driving strengths, temperatures, and magnitudes of exciton-phonon couplings that are being explored experimentally have not so far been captured by a single, intuitive, quantum master equation, making the interpretation of results potentially difficult.

Existing techniques for calculating the density matrix dynamics of an open quantum system include methods which can numerically converge to give exact results; one example is the quasi-adiabatic propagator path integral (QUAPI).~\cite{makri95} However, such numerical methods can be very computationally expensive in certain parameter regimes, and typically require a complementary approach in order to fully appreciate the complex 
dynamical behaviour they predict.  
In contrast, quantum master equations 
naturally relate the evolution of density matrix elements with environmental parameters in an intuitive way, and are often less computationally expensive. However, the derivation of a practical master equation always involves an approximation, and a new equation is typically derived for each distinct parameter regime. In the case of driven quantum dots, popular choices have included a weak coupling approach,~\cite{ramsay10,nazir08,machnikowski04} whose validity relies on weak exciton-phonon coupling; and a polaron transform method~\cite{wilson02,mccutcheon10_2, roy11} which works for stronger coupling, but which 
in turn introduces conditions on the driving strength.~\cite{mccutcheon10_2} 

In this work, we develop a physically-motivated, versatile and efficient master equation approach to describe the dynamics of a continuously-driven QD in contact with a phonon environment, through a combination of a variationally-optimized unitary transformation and the time-local projection operator technique. In those limits appropriate for either a weak-coupling or polaron treatment we find that our master equation yields similar results, but it is 
also able to capture the excitonic dynamics over a much wider range of parameters where such simpler treatments fail.
The accuracy of our variational master equation is verified by its excellent agreement with numerically converged QUAPI calculations. 

{\it Variational transformation} - 
As is common,~\cite{ramsay10, ramsay10_2,nazir08, mccutcheon10_2,vagov2007,krugel05, machnikowski04, forstner03} we model our QD as a two-level system with ground-state $|0\rangle$ and excited (single-exciton) state $|X\rangle$, separated by an energy of $\hbar\omega_0$. The dot is driven by a laser of frequency $\omega_l$, with time-independent Rabi frequency $\Omega$, and is coupled to an acoustic phonon environment 
represented by a harmonic oscillator bath of frequencies $\omega_{\bf k}$, coupling strengths $g_{\bf k}$, and creation operators $b_{\bf k}^{\dagger}$. In a frame rotating at frequency $\w_0$, and after a rotating-wave approximation on the driving term, the total Hamiltonian reads (setting $\hbar=1$)
\begin{equation}
\label{HRWA}
H=\Big[\delta+\sum_{\bf k}g_{\bf k}(b_{\bf k}^{\dagger}+b_{\bf k})\Big]|X\rangle\langle X|+\frac{\Omega}{2}\sigma_x
+\sum_{\bf k}\omega_{\bf k}b_{\bf k}^{\dagger}b_{\bf k},
\end{equation}
where $\delta=\omega_0-\omega_l$ is the detuning of the laser from the excitonic transition energy, $(\Omega/2)\sigma_x=(\Omega/2)(|0\rangle\langle X|+|X\rangle\langle 0|)$ drives coherent excitonic oscillations, and an irrelevant term proportional to the identity 
has been neglected. 

To achieve our aim of constructing a master equation 
that is valid over as wide a range of parameters as possible, we employ a technique similar to that originally developed by Silbey and Harris to study the ground state of the spin-boson model.~\cite{silbey84} We thus proceed by applying a unitary 
transformation to $H$ that displaces the bath oscillators 
according to the state of the QD excitonic system, by an amount that is determined by a set $\{f_\kv\}$ of variational parameters which we shall find through free energy minimization arguments. Given the particular form of Eq.~(\ref{HRWA}), we define 
$H_{\rm V }=\e^{V}H\e^{-V}$, where $\e^{\pm V}=\ketbra{0}{0}+\ketbra{X}{X}\prod_\kv D(\pm\alpha_\kv)$, with $D(\pm\alpha_\kv)=\exp[\pm\alpha_\kv(b_\kv^{\dagger}-b_\kv)]$ a displacement operator and 
$\alpha_\kv=f_\kv/\w_\kv$ assumed to be real. Notice that if $f_\kv$ is fixed equal to $g_\kv$ (for all modes) the transformation is equivalent to the full polaron displacement,~\cite{mahan, mccutcheon10_2, wilson02,roy11} but here we explicitly give ourselves more freedom in choosing each $f_\kv$ such that we obtain 
more accurate dynamics. The idea is that by moving into a representation in which 
the displacement of the bath in response to the system is already accounted for, and by allowing some freedom to optimise the transformation through the set $\{f_\kv\}$, it may be possible to identify a small interaction term in $H_{\rm V}$ (to be treated as a perturbation) even if there is no obvious small term in the original Hamiltonian $H$.

After transformation, the Hamiltonian reads 
\begin{equation}\label{originaltransformedH}
H_{\rm V}=\frac{R}{2}\openone +\frac{\epsilon}{2}\sigma_z+H_B+\frac{\Omega}{2}\left(\sigma_+B_++\sigma_-B_-\right)+\ketbra{X}{X}B_z,
\end{equation}
where $H_B=\sum_\kv\w_\kv b_\kv^{\dagger}b_\kv$, and $\sigma_z=\ketbra{X}{X}-\ketbra{0}{0}$. The detuning is now shifted by an amount dependent upon the variational parameters: $\epsilon=\delta+R$, where 
$R=\sum_\kv \w_\kv^{-1}f_\kv(f_\kv-2g_\kv)$, which in the limit $f_\kv\rightarrow g_\kv$ (full polaron displacement) becomes the polaron shift, $R\rightarrow-\sum_\kv \w_\kv^{-1}g_\kv^2$. Note that since the term proportional to the identity in Eq.~({\ref{originaltransformedH}}) also depends on the variational parameters, it cannot be neglected in the free energy minimisation step. The bath operators are now given by $B_z=\sum_\kv(g_\kv-f_\kv)(b_\kv^{\dagger}+b_\kv)$ and $B_{\pm}=\prod_\kv D(\pm\alpha_\kv)$. It is straightforward to show that $\langle B_{z}\rangle_{H_B}={\rm Tr} \{B_{z}\exp[-\beta H_B]\}/\mathrm{Tr}\{\exp[-\beta H_B]\}=0$, for inverse temperature $\beta=1/k_B T$, but the same is not necessarily true for $B_{\pm}$. We thus define $B=\langle B_{\pm}\rangle_{H_B}$, subtract this factor from the relevant interaction terms in Eq.~(\ref{originaltransformedH}), and add them instead to the system Hamiltonian. We may then split the transformed Hamiltonian as $H_{\rm V}=H_0+H_I$ with the free Hamiltonian chosen to be $H_0=H_S+H_B$, where
\beq
H_S=\frac{R}{2}\openone+\frac{\epsilon}{2}\sigma_z+\frac{\Omega_r}{2}\sigma_x,
\label{free_hamiltonian}
\eeq
with $\Omega_r \equiv B\Omega$, while the interaction Hamiltonian becomes
\beq
H_I=\frac{\Omega}{2}\Big(B_x\sigma_x+B_y\sigma_y\Big)+\ketbra{X}{X}B_z,
\label{H_I}
\eeq
written in terms of the Hermitian combinations $B_x=(1/2)(B_++B_--2B)$ and $B_y=(1/2i)(B_--B_+)$.

Importantly, in splitting $H_{\rm V}$ in the way outlined above, we have removed the term $(\Omega_r/2)\sigma_x$ from $H_I$ to include it instead in $H_S$. This has accomplished two things: (i) it ensures that $\langle H_I\rangle_{H_0}=0$, which simplifies the form of the master equation to be derived below; (ii) more importantly, the term now appears as a {\it bath-renormalised} 
driving in Eq.~({\ref{free_hamiltonian}}) and is therefore treated non-perturbatively in the subsequent formalism. This ensures that our theory can capture coherent exciton dynamics at the correct phonon-dressed energy scale $\Omega_r$. For a stationary bath state $\rho_B=\exp[-\beta H_B]/\mathrm{Tr}\{\exp[-\beta H_B]\}$ we find the renormalisation factor
$B=\exp[-(1/2)\sum_\kv \alpha_\kv^2 \coth(\beta\w_\kv/2)]$.

{\it Free energy minimisation} - 
We find the set of variational parameters $\{f_\kv\}$ by imposing that the free energy associated with the transformed Hamiltonian be minimised.~\cite{silbey84} In doing so, we are attempting to find the best approximate diagonalization of $H$ given the restricted form of the transformation $e^VHe^{-V}$, and thus expect contributions  
from the interaction Hamiltonian $H_I$ to be minimised as well. In effect, the variational parameters allow us to tune which parts of the total Hamiltonian are treated exactly, and which are treated in a perturbative manner. 

To proceed, we compute the Feynman-Bogoliubov upper bound on the free energy, given by~\cite{silbey84}
\beq
A_B=-\frac{1}{\beta}\ln\bigl(\mathrm{Tr}\{\e^{-\beta H_{0}}\}\bigr)+\langle H_{I} \rangle_{H_{0}}+\mathcal{O}( \langle H_{I}^2 \rangle_{H_{0}}),
\label{A_B}
\eeq
and related to the true free energy, $A$, via the inequality $A_B\geq A$. By construction, the second term in Eq.~({\ref{A_B}}) is equal to zero. Neglecting 
higher order terms then leads to the minimization condition,
\begin{align}
f_\kv=\frac{g_\kv(1-\frac{\epsilon}{\eta}\tanh(\beta\eta/2))}
{1-\frac{\epsilon}{\eta}\tanh(\beta\eta/2)\Big(1-\frac{\Omega_r^2}{2\epsilon\w_\kv}\coth(\beta \w_\kv/2)\Big)},
\label{minimisation_condition}
\end{align}
which self-consistently determines the set of variational parameters $\{f_\kv\}$. Here, $\eta=\sqrt{\epsilon^2+\Omega_r^2}$ is the system eigenstate energy splitting in the variational frame. 
It is now insightful to look at the form of $f_\kv$ in two opposite limits (assuming, for simplicity, that we can set the detuning $\epsilon$ to be zero): (i) when $\Omega<<\omega_\kv$ we find $f_\kv\approx g_\kv$, and we recover the full polaron transformation. In this situation, the driving is weak enough that the bath oscillators can follow the excitonic motion and become fully displaced when the system is in its excited state, as determined by the form of coupling in $H$; (ii) on the other hand, when $\Omega>>\omega_\kv$, we find that $f_\kv$ becomes very small, so that there is barely any displacement due to the transformation. In this case, the excitonic oscillations are too fast for the relevant bath modes to follow, and 
their displacements 
are thus suppressed. We shall see that this has important physical consequences in the context of driven QDs as it gives rise to a reduction in phonon-induced damping at large driving strengths within the variational theory, as originally predicted using a Feynman integral approach.~\cite{vagov2007}

Writing $f_\kv=g_\kv F(\w_\kv)$ and defining the spectral density $J(\w)=\sum_\kv|g_\kv|^2\delta(\w-\w_\kv)$, we find the following self-consistent forms for $B$ and $R$ in the continuum limit:
\begin{align}
B=&\exp\Big[-(1/2)\int_0^{\infty}J(\w)F(\w)^2\frac{\coth(\beta\w/2)\mathrm{d}\w}{\w^2}\Big];
\label{Bav}\\
R=&\int_0^{\infty}J(\w)F(\w)\w^{-1}(F(\w)-2)\mathrm{d}\w.
\label{R_integral}
\end{align}
For deformation potential coupling of the exciton state to acoustic phonons 
we can take a super-Ohmic spectral density of the form $J(\w)=\alpha\,\w^3\e^{-(\w/\w_c)^2}$,~\cite{Krummheuer02,ramsay10, ramsay10_2} where $\alpha$ captures the strength of the exciton-phonon coupling and $\w_c$ provides a natural high frequency cut-off, which is proportional to the inverse of the carrier localization length in the QD.~\cite{Krummheuer02} Throughout this paper we use the experimentally determined values of $\alpha=0.027$~ps$^2$ and $\omega_c=2.2$~ps$^{-1}$.~\cite{ramsay10_2} In general, for given values of $\alpha$, $\w_c$, $\Omega$, $\delta$ and $\beta$, Eqs.~({\ref{Bav}}) and ({\ref{R_integral}}) must be solved simultaneously, which is straightforward numerically.

{\it Master equation} - 
To describe the dynamics of the driven QD under the influence of the phonon environment, we employ the time-convolutionless projection operator technique~\cite{b+p} to obtain a rigorous 
master equation within the variationally-transformed representation. We consider the QD to be initialized in its ground state, for which the appropriate state of the bath is a thermal equilibrium state 
(since there is no system-bath coupling through the QD ground-state). We therefore write the initial total density operator as 
$\chi(0)=\ketbra{0}{0}\otimes\rho_B$, from which we find $\e^{V}\chi(0)\e^{-V}=\chi(0)$; our initial state is unaffected by the transformation. By choosing $\rho_B$ as our reference state and truncating 
at second-order in $H_I$, we thus find a homogeneous interaction picture master equation 
\beq
\frac{\mathrm{d}\tilde{\rho}_V}{\mathrm{d} t}=-\int_0^{t}\mathrm{d}s\mathrm{Tr}_B[\tilde{H}_I(t),[\tilde{H}_I(s),\tilde{\rho}_V(t)\rho_B]],
\label{me_1}
\eeq
where $\rho_V(t)=\mathrm{Tr}_B(\e^V\chi(t)\e^{-V})$ 
describes the QD excitonic degrees of freedom in the variational frame, 
with $\mathrm{Tr}_B$ denoting a trace over the environment (phonon) degrees of freedom, and $\tilde{O}(t)=e^{i H_0 t}Oe^{-i H_0 t}$.  Using Eq.~({\ref{H_I}}) and returning to the Schr\"{o}dinger picture, we then obtain
\begin{align}
\frac{\mathrm{d}{\rho}_V}{\mathrm{d} t}&=-i[H_S,\rho_V(t)]\nonumber\\
&-{\textstyle{\frac{1}{2}}}\sum_{ij}\sum_{\w}\gamma_{ij}(\w,t)[A_i,A_j(\w)\rho_V(t)-\rho_V(t)A_j^{\dagger}(\w)]\nonumber\\
&-i\sum_{ij}\sum_{\w}S_{ij}(\w,t)[A_i,A_j(\w)\rho_V(t)+\rho_V(t)A_j^{\dagger}(\w)],
\label{me_2}
\end{align}
where $\{i, j\}\in\{1,2,3\}$ and $\omega\in\{0,\pm\eta\}$. We define $A_1=\sigma_x$, $A_2=\sigma_y$ and $A_3=(1/2)(I+\sigma_z)$, while $A_1(0)=\sin 2\theta(\ketbra{+}{+}-\ketbra{-}{-})$, $A_1(\eta)=\cos 2\theta\ketbra{-}{+}$, 
$A_2(0)=0$, $A_2(\eta)=i \ketbra{-}{+}$, $A_3(0)=\cos^2\theta\ketbra{+}{+}+\sin^2\theta\ketbra{-}{-}$ and 
$A_3(\eta)=-\sin\theta\cos\theta\ketbra{-}{+}$, defined in terms of the eigenstates of $H_S$, satisfying 
$H_S\ket{\pm}=(1/2)(R\pm\eta)\ket{\pm}$. In all cases $A_i(\w)=A_i^{\dagger}(-\w)$. The angle 
$\theta=(1/2)\arctan(\Omega_r/\epsilon)$ characterises the tilt of the system eigenstates away from the $x$-axis in the variational frame.

\begin{figure}[!t]
\includegraphics[width=0.48\textwidth]{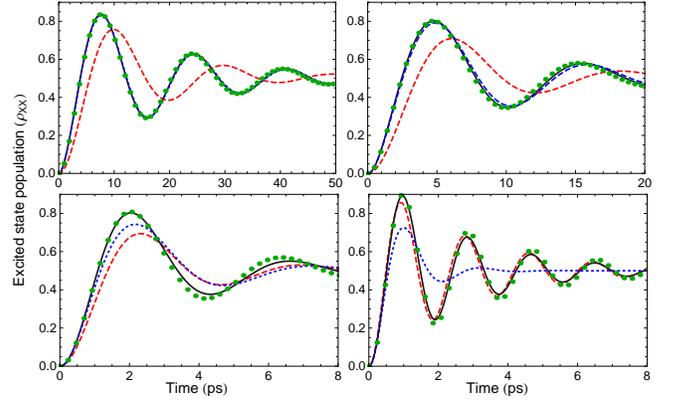}
\caption{Exciton population dynamics 
calculated from the variational master equation (grey solid curves), weak-coupling theory (red dashed curves), polaron theory (blue dotted curves), and the QUAPI (green points), for $T=50$~K. Here, $\Omega=\pi/6$~ps$^{-1}$ (top left), $\Omega=\pi/4$~ps$^{-1}$ (top right), $\Omega=\pi/2$~ps$^{-1}$ (bottom left) and $\Omega=\pi$~ps$^{-1}$ (bottom right). Spectral density parameters: $\alpha=0.027$~ps$^2$, $\omega_c=2.2$~ps$^{-1}$.}
\label{T50dynamics}
\end{figure}

Eq.~({\ref{me_2}}) contains the 
rates and energy shifts 
$\gamma_{ij}(\w,t)=2\mathrm{Re}[K_{ij}(\w,t)]$ and $S_{ij}(\w,t)=\mathrm{Im}[K_{ij}(\w,t)]$, respectively, defined in terms of the response functions
\beq\label{bathresponse}
K_{ij}(\w,t)=\int_0^t \Lambda_{ij}(\tau)\e^{i\w t}\mathrm{d}\tau,
\eeq
which themselves depend on the bath correlation functions $\Lambda_{ij}(\tau)=\mathrm{Tr}(\tilde{B}_i(\tau)\tilde{B}_j(0)\rho_B)$. Here we label $B_1=(\Omega/2)B_x$, $B_2=(\Omega/2)B_y$, and $B_3=B_z$. The bath correlation functions are most easily calculated by utilising the coherent state representation of the bath density operator.~\cite{glauber63} 
We find $\Lambda_{11}(\tau)=(\Omega_r^2/8)(\e^{\phi(\tau)}+\e^{-\phi(\tau)}-2)$ and $\Lambda_{22}(\tau)=(\Omega_r^2/8)(\e^{\phi(\tau)}-\e^{-\phi(\tau)})$, 
with phonon propagator 
\begin{align}
\phi(\tau)=&\int_0^{\infty}\mathrm{d}\w\frac{J(\w)}{\w^2}F(\w)^2G_+(\tau),
\intertext{defined in terms of $G_{\pm}(\tau)=(n(\w)+1)e^{-i\w\tau}\pm n(\w)\e^{i\w\tau}$, with 
$n(\w)=(e^{\beta \w}-1)^{-1}$ the occupation number, while}
\Lambda_{33}(\tau)=&\int_0^{\infty}\mathrm{d}\w J(\w)(1-F(\w))^2G_+(\tau),\\
\Lambda_{32}(\tau)=&\frac{\Omega_r}{2}\int_0^{\infty}\mathrm{d}\w\frac{J(\w)}{\w}F(\w)(1-F(\w))iG_-(\tau),
\end{align}
with $\Lambda_{32}(\tau)=-\Lambda_{23}(\tau)$, and $\Lambda_{12}(\tau)=\Lambda_{21}(\tau)=\Lambda_{13}(\tau)=\Lambda_{31}(\tau)=0$.

\begin{figure}[!t]
\includegraphics[width=0.49\textwidth]{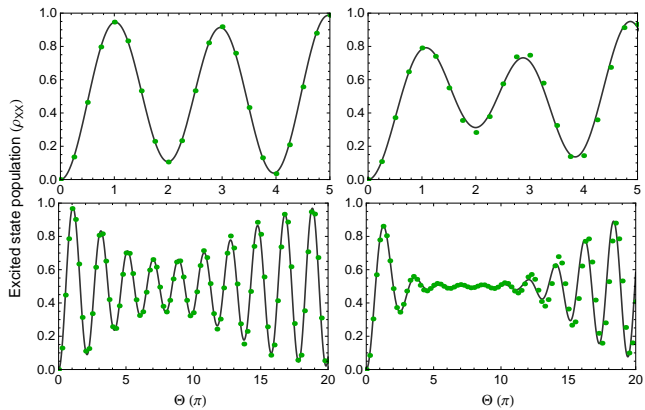}
\caption{Exciton population as a function of pulse area ($\Theta=\Omega\tau$) for pulse durations of $\tau=2.5$~ps (top) and $\tau=10$~ps (bottom), and 
for $T=10~K$ (left) and $T=50~K$ (right). Variational master equations results are shown by solid lines with the corresponding QUAPI 
results shown with green points. Spectral density parameters: $\alpha=0.027$~ps$^2$, $\omega_c=2.2$~ps$^{-1}$.} 
\label{T50pulse}
\end{figure}

{\it Dynamics and comparison with the path integral approach} - 
In order to clearly demonstrate the superiority of the variational master equation over either the weak-coupling or polaron forms, we shall now compare 
its predictions with numerically accurate results calculated using the well-established QUAPI technique.~\cite{makri95} In all plots shown we have checked for numerical convergence of the QUAPI results.

In Fig.~\ref{T50dynamics} we plot the excited state population dynamics ($\rho_{\rm XX}$) of the QD  as a function of time,~\footnote{In all plots we set the laser to be on resonance with the full polaron-shifted excitonic transition energy, $\omega_l=\omega_0+R$.} generated from the present variational master equation, the polaron master equation of Ref.~\onlinecite{mccutcheon10_2}, the weak-coupling master equation of Ref.~\onlinecite{nazir08}, and our QUAPI benchmarks. For the realistic parameters chosen here, the variational theory is in excellent agreement with the QUAPI results across the full range of driving strengths. Importantly, neither the polaron nor weak-coupling theories can match the QUAPI results across this entire range. The polaron theory fails, as anticipated in Ref.~\onlinecite{mccutcheon10_2}, at strong driving, $\Omega>\omega_c$. Perhaps surprisingly, the weak-coupling theory works well when driving beyond the cut-off frequency, even though it fails at weaker driving strengths due to the relatively large temperature of $T=50$~K. In the latter regime, phonon-induced driving renormalisation and multi-phonon effects, which the weak-coupling theory cannot capture, are especially important.

In order to interpret the behaviour seen as $\Omega$ is varied in Fig.~\ref{T50dynamics} it is instructive to consider the appropriate 
limiting forms of the master equation~({\ref{me_2}}), determined by the bath correlation functions $\Lambda_{ij}(\tau)$. 
For weak driving conditions (though not weak system-bath coupling), we saw earlier that the variational transformation reduces 
approximately to performing the full polaron transformation on $H$, which corresponds to the limit $F(\w)\rightarrow1$ (for all modes). 
In this limit, only $\Lambda_{11}(\tau)$ and $\Lambda_{22}(\tau)$ survive and we recover the polaron master equation presented in 
Ref.~\onlinecite{mccutcheon10_2}, where now the perturbative expansion corresponds to one in the phonon-dressed driving term. On the other hand, when driving beyond the cutoff frequency, $\Omega>\omega_c$, essentially all relevant bath oscillators become sluggish and they are now unable to dress the system, so the full polaron transformation is no 
longer appropriate. As outlined before, the correct limit is now $F(\w)\rightarrow0$, such that effectively no transformation is applied to the Hamiltonian and our perturbative expansion now corresponds to one in the original exciton-phonon coupling strength. Setting $F(\w)=0$ we find that only $\Lambda_{33}(\tau)$ survives 
and our master equation reduces to the weak-coupling form presented in Ref.~\onlinecite{nazir08}.  
In general, however, the minimisation condition given by Eq.~({\ref{minimisation_condition}}) corresponds to neither of these limiting cases, and $F(\w)$ acts to optimise the transformation for each phonon mode depending on the specific Hamiltonian parameters. 
The variational master equation~({\ref{me_2}}) therefore describes an intricate, yet physically motivated mixture of the weak-coupling and polaron approximations.

To further confirm the versatility of the variational master equation, 
in Fig.~\ref{T50pulse} we plot the common experimental scenario of excitonic population measured as a function of pulse area, $\Theta=\Omega\tau$,~\cite{ramsay10, ramsay10_2} for various pulse-widths $\tau$. Again, for all cases the variational theory matches the QUAPI results very closely, and in particular correctly predicts a recovery of Rabi oscillation amplitude for larger pulse areas.~\cite{vagov2007} 

{\it Summary} - 
We have derived a variationally-optimized master equation for a two-level system interacting with a bosonic environment, and applied it to study the dynamics of phonon-damped exciton Rabi rotations in a laser-driven QD. By comparison with the numerically accurate QUAPI approach, we have shown that our variational master equation can give quantitatively accurate results in difficult parameter regimes, where simpler master equation treatments fail. More generally, our results imply that, for super-Ohmic environments at least, the variational master equation can give accurate results for the dynamics of dissipative systems in distinct parameter regimes (i.e. where the bath displacement does or does not play a strong role) and can reliably interpolate between them. Our approach is attractive not only for its relative simplicity, but also for the physical insight it 
gives to the system under study. In the present context, we 
see how the reduced damping observed 
using the Feynman integral at large pulse areas~\cite{vagov2007} naturally arises in the variational theory due to the sluggish nature of the bath oscillators in this regime.

{\it Acknowledgments.} 
A. N. thanks Imperial College, B.W.L. thanks the Royal Society, A.N., D.P.S.M. and E. M. G. thank the EPSRC, and N.S.D. thanks the Clarendon Fund and the NSERC/CRSNG of/du Canada  for financial support.

\end{document}